\documentclass[showpacs,amsmath, amssymb, amsthm, aps, jmp, thmsb, twocolumn]{revtex4}
\usepackage{color,amsbsy,amssymb,latexsym,amsfonts, amsmath,cancel}
\usepackage{mathrsfs}
\usepackage{graphicx}

\usepackage{subfigure}

\newcommand{\bel}[1]{\begin{equation}\label{#1}}                     
\newcommand{\bal}[1]{\begin{eqnarray}\label{#1}}   
\newcommand{\be}{\begin{equation}}               
\newcommand{\ba}{\begin{eqnarray}}           
\newcommand{\ee}{\end{equation}}
\newcommand{\ea}{\end{eqnarray}}

\newcommand{\bea}{\begin{equation}}
\newcommand{\eea}{\end{equation}}
\begin{document}
\title{Quantum Trajectories based on the Weak Value}

\author{Takuya Mori$^1$}
\email[Email:]{takumori@post.kek.jp}
\author{Izumi Tsutsui$^{1, 2}$}
\email[Email:]{izumi.tsutsui@kek.jp}

\affiliation{$^1$Department of Physics, University of Tokyo, 7-3-1 Hongo, Bunkyo-ku, Tokyo 113-0033, Japan\\
$^2$Theory Center, Institute of Particle and Nuclear Studies,
High Energy Accelerator Research Organization (KEK), 1-1 Oho, Tsukuba, Ibaraki 305-0801, Japan
}

\begin{abstract}%
The notion of the trajectory of an individual particle
is strictly inhibited in quantum mechanics because of the uncertainty principle.  Nonetheless, the weak value, which has been proposed as a novel and measurable quantity definable to any quantum observable, can offer a possible description of trajectory on account of its statistical nature.
In this paper, we explore the physical significance provided by this ``weak trajectory''  by considering various situations where interference takes place simultaneously with the observation of particles, that is, in prototypical quantum situations for which no classical treatment is available.  These include the double slit experiment and Lloyd's mirror, where in the former case it is argued that the real part of the weak trajectory describes an average over the possible classical trajectories involved in the process, 
and that the imaginary part is related to the variation of interference.  
It is shown that 
this average interpretation of the weak trajectory holds universally under the complex probability defined from the given transition process.  These features remain essentially unaltered in the case of Lloyd's mirror where interference occurs with a single slit.
\end{abstract}

\maketitle

\section{Introduction}

In quantum mechanics, we all know that one cannot associate a trajectory to a particle on an observational basis 
because, as dictated by the uncertainty principle, any measurement of the position disturbs the momentum of the particle, rendering the successive determination of the position within a certain limit of accuracy impossible.   
A typical situation illustrating this is found in the double slit experiment, where the assumption of trajectory--the particle goes through either of the two slits--contradicts the interference pattern actually observed on the screen.  

In this respect, the weak value, which was proposed earlier by Aharonov {\it et al.}~\cite{Aharonov} as a novel physical quantity obtainable with a minimal disturbance, offers an interesting possibility.  
Namely, under the given process of transition from an initial state, called the ``pre-selected state,'' $\vert \phi\rangle$, and a final state called the ``post-selected state,'' $\vert \psi\rangle$, for which the process is assumed not to be forbidden: $\langle \psi |\phi\rangle \ne 0$, 
one can measure an observable $A$ by the standard von Neumann measurement to obtain the {\it weak value},
\begin{eqnarray}
A_w =\frac{\langle \psi| A |\phi\rangle}{\langle \psi |\phi\rangle},
\label{wv}
\end{eqnarray}
in the weak (vanishing) limit of the measurement interaction.  As such, one may expect that the weak value $A_w$ has something to say about the physical being of the observable $A$ peculiar to the process, revealing its raw features that are accessible only under the undisturbed condition.
In the same vein, one may also contemplate that the weak value, albeit being complex, has an ``element of reality'' \cite{Vaidman}  analogous to that argued in the Einstein-Podolsky-Rosen paper \cite{EPR}.

Recently, Kocsis {\it et al.}~demonstrated experimentally that particle trajectory in the double slit experiment can be reconstructed for an ensemble of particles by combining the weak value of the momentum and the determination of the position in a systematic way \cite{Kocsis}.   The emergent trajectory agrees with that of the de Broglie-Bohm theory, confirming the prediction made in \cite{QuantumPotential,Wiseman}.   This is an enlightening result illuminating the physical meaning of the weak value, but in view of the indirect nature of the systematic treatment of the data, one may wonder if there are  other ways to define the trajectory based on the weak value.  Indeed, an alternative and more direct trajectory can be delineated from the time-dependent weak value \cite{Tanaka} with an iterative procedure of measurements \cite{Matzkin}.  This can be regarded as the dynamical version of the weak value that emerges naturally in the context of the time-symmetric formulation of quantum mechanics \cite{ABL}.   

In the present paper, we consider the dynamical trajectory defined directly in this context, called the ``weak trajectory'' for short, 
and thereby examine if it allows for an intuitive picture from the physical point of view.  When the given process consists of the pre- and post-selections given either by position eigenstates or by well-localized (Gaussian) states, the weak trajectory forms a curve with the two ends specified by the selections.   The interest in this case is then to see how the trajectory deviates from the classical one, which has been analyzed earlier in \cite{Tanaka, AharonovAve, Matzkin}.  In contrast, in our study we are interested in situations where the selections are performed not by those (semi)classical states but by genuinely quantum states, that is, by superposed states for which no particular position is assignable to the particle, at one of the ends at least.  Specifically, we consider a number of examples including the double slit experiment and its extensions to multiple slit and further to the continuous case, and also Lloyd's mirror for which no unique classical trajectory is available.   These examples offer us a reasonably good testing ground for the validity of the weak trajectory.

Another question we address is the generic complex-valuedness of the weak trajectory implied in the definition of the weak value (\ref{wv}).  Through the examples we study, we find that, albeit with complex probability for the general case,
the weak trajectory admits interpretation as an average over the possible distinct trajectories involved in the transition process, while the imaginary part is related to the rate of variation in the interference observed.   The condition for the weak trajectory to become entirely real will be discussed in some detail.   
Interestingly, we shall see that the average nature of the weak
trajectory can be made into an individual one while preserving the interference, if we install a device to provide the which-path information by measuring the spin of the particle at the time of the post-selection \cite{Mori}.
In the more general case, including the multiple slit case, we find that
the interpretation of the weak value can be maintained if we adopt the extended notion of complex probability assigned to the process of transition mentioned earlier \cite{Hosoya, Morita}.  
Finally, in the example of Lloyd's mirror in which interference with a single slit is realized, the weak trajectory is mostly seen to yield a smoothed average of the two classical trajectories appearing in the process.   

The plan of the paper is as follows.  After providing some preliminaries in Sect.\,II necessary for our later discussions, we discuss in Sect.\,III the weak trajectory in the double slit experiment.   Then we go on to consider the triple slit case in Sect.\,IV and, further, the multiple slit case in Sect.\,V together with the reality condition of the trajectory.  
Section VI is devoted to the question of how to obtain the which-path information that is available in these experiments.  We then study the completely general case of pre-selection in Sect.\,VII, where the momentum eigenstate is seen to yield a classical picture for the weak trajectory.  Lloyd's mirror is treated in Sect.\,VIII before Sect.\,IX provides our conclusion and discussions.

\section{Preliminaries}

Our primary concern in this paper is the question of how the weak value develops in time, and for this we consider the time-dependent weak value,
\begin{eqnarray}
A_w(t) :=\frac{\langle \psi|U(T-t)\, A\, U(t)|\phi\rangle}{\langle \psi|U(T)|\phi\rangle},
\label{tdwv}
\end{eqnarray}
which represents the outcome of the weak measurement of an observable $A$ performed at time $t \in [0, T]$.
Here, the measurement is made under the premise that the system undergoes the transition process starting from 
the pre-selected state $|\phi\rangle$ at $t = 0$ and ending with the post-selected 
state $|\psi\rangle$ at $t = T$.   The unitary operator $U(t)$ describes the time development of the system from the initial
$t = 0 $ to an arbitrary intermediate time $0 < t < T$ when the value of the observable $A$ is measured weakly, and likewise $U(T-t)$
describes the time development from that moment to the final $t = T$. 
From the conceptual viewpoint advocated in the time-symmetric formulation of quantum mechanics \cite{ABL}, the formula (\ref{tdwv}) is just the standard weak value (\ref{wv}) evaluated at time $t$ under
the pre-selected state which evolves {\it forward} in time to $U(t)|\phi\rangle$ and the post-selected state which evolves {\it backward} in time to $U(t-T)|\psi\rangle$.  

Note that the quantity $A_w(t)$ may be regarded as an extension of the expectation value, since it reduces to the conventional expectation value of the Heisenberg operator $A(t) = U(t)^{-1} A U(t)$ in the particular case where we have  $|\psi\rangle = U(T)|\phi\rangle$, that is, the post-selected state happens to be the time-developed pre-selected state.
As such, for the unitary time development $U(t) = \exp({-{i} H t}/\hbar)$ governed by the Hamiltonian $H$, the weak value $A_w(t)$ obeys the equation
\begin{eqnarray}
\frac{d}{dt}A_w(t)
&=&-\frac{{i}}{\hbar}\frac{\langle \psi|U(T-t)\, [A,H]\,U(t)|\phi\rangle}{\langle \psi|U(T)|\phi\rangle}
\nonumber\\
&=&-\frac{{i}}{\hbar}[A,H]_w(t),
\label{Ehrenfest}
\end{eqnarray}
which is analogous to one stipulated by the Ehrenfest theorem for the expectation value, despite that $A_w(t)$ is complex in general.

In what follows we consider the system of a particle of mass $m$ under the non-relativistic Hamiltonian $H=\frac{p^2}{2m}+V(x)$.  If, in particular, the particle is free ($V(x) = 0)$, by putting $p$ and $x$ for the observable $A$ in (\ref{Ehrenfest}) we find that the momentum weak value $p_w$ and the position weak value $x_w$ obey
\begin{eqnarray}
&&\frac{d}{dt}p_w(t)=0,
\label{EhrenfestDoubleMome}\\
&&\frac{d}{dt}x_w(t)=\frac{1}{m} p_w(t).
\label{EhrenfestDoublePosi}
\end{eqnarray}
It then follows from (\ref{EhrenfestDoubleMome}) and (\ref{EhrenfestDoublePosi}) that 
$x_w(t)$ is a linear (complex) function of $t$.   This implies that $x_w(t)$  is a real function during the entire interval $t \in [0, T]$ if and only if both the initial and final values $x_w(0)$ and $x_w(T)$ are real.  
This occurs, for instance, when both of the two selections are made by position eigenstates, $|\phi\rangle = |x_i\rangle$ and $|\psi\rangle = |x_f\rangle$, 
in which case we have the endpoints, $x_w(0) = x_i$ and $x_w(T) = x_f$, and accordingly the position weak value, or the {\it weak trajectory},   
\begin{eqnarray}
x_w(t) =\frac{\langle x_f|U(T-t)\,x\,U(t)| x_i\rangle}{\langle x_f|U(T)| x_i\rangle},
\label{wtsingle1}
\end{eqnarray}
coincides with the classical trajectory $x_w(t) = x_{cl}(t)$ given by
\begin{eqnarray}
x_{cl}(t) :=\frac{(x_f - x_i)t +  x_iT}{T}.
\label{wtsingle}
\end{eqnarray}

We thus learn that, at least in the simple situation of a free particle residing at a particular location at the ends $t = 0$ and $t = T$, the particle trajectory in quantum mechanics viewed in terms of the weak value 
provides a reasonable intuitive picture of the location of the particle during the period $[0, T]$: it follows precisely the classical trajectory.  
This reassuring feature of the weak trajectory will, of course, no longer be valid when the particle is not free, or when the particle does not reside at a particular location at the ends.  These two cases where the simple outcome cannot be expected possess distinct characteristics of their own.

In the former case, despite that the weak trajectory will give a different path from the classical one, it still yields some unique path which may be given a physical significance in one way or another.  The latter case, where the particle can reside at more than one point, occurs if we choose pre- or post-selections by a superposition of more than one position eigenstates.  Obviously, this poses a more serious problem for the interpretation of the weak trajectory, because of the nonlocality inherent to the generic quantum states.
Starting with the next section, we shall provide a case study of this latter case to examine what happens when the pre-selection is made nonlocal, starting with the typical example offered by the double slit experiment and then generalizing it gradually.  In the last example, we touch upon the case of Lloyd's mirror, where the element of the former case is also involved.

\section{The Double Slit Experiment}

\begin{figure}[t]
\begin{center}
\includegraphics[width=5.8cm]{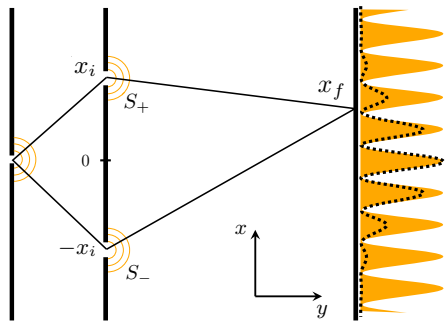}
\caption{Our simplified double slit (gedanken) experiment.  
The orange filled curve describes the transition probability in our idealized situation where the slits $S_\pm$ are treated as point-like, whereas the dotted curve describes the transition probability in a more realistic situation where the slits $S_\pm$ are of a finite size and the particle distribution is Gaussian. 
}
\label{QuantumInterference}
\end{center}
\end{figure}

As a first attempt at examining the weak trajectory in a nontrivial case, let us consider the case in which the pre-selection is made by the 
the superposition of the two position eigenstates, $|x_i\rangle$ and $|-x_i\rangle$, {\it i.e.}, 
\begin{eqnarray}
|\phi\rangle=\frac{1}{\sqrt{2}}\left(|x_i\rangle+|-x_i\rangle\right)
\end{eqnarray}
whereas the post-selection remains as the position eigenstate $|\psi\rangle = |x_f\rangle$.  Clearly, this offers a simplified version of 
the double slit (gedanken) experiment in which a Gaussian distribution around the slits is replaced by the pre-selected state (see FIG.\,\ref{QuantumInterference}).   For simplicity of the analysis, in the following we consider only the degrees of freedom $(x, p)$ 
which are parallel to the screen, ignoring those perpendicular to the screen which are inessential for our argument. 

\begin{figure}[t]
\begin{center}
\includegraphics[width=6cm]{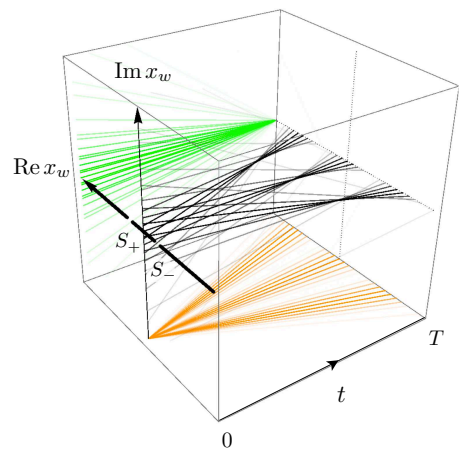}
\caption{ 
Weak trajectories $x_w(t)$ in the complex plane for various different post-selections.  The lines are plotted with density proportional to the transition probability $ |\langle\psi|U(T)|\phi\rangle|^2$.   The real and imaginary parts are depicted in orange and green lines and projected on the bottom and the left-back planes, respectively.
}
\label{InterferenceDiagonal}
\end{center}
\end{figure}
\hspace{5mm}

Assuming that the particle is governed by the free Hamiltonian, one has the Feynman kernel for the transition amplitude,
\begin{eqnarray}  
\langle x_f|U(T)|x_i\rangle = \left({{m}\over {2\pi i \hbar T}}\right)^{1/2}
\exp \left[ {{i}\over{\hbar}} {{m}\over{2}} {{(x_f - x_i)^2}\over{T}} \right],
\label{transamp}
\end{eqnarray}
from which the transition probability showing the interference pattern on the screen follows immediately,
\begin{eqnarray}
\vert \langle x_f|U(T) \vert \phi \rangle\vert^2 
= {{m}\over {2\pi \hbar T}}
\left\{1 + \cos \left(\frac{m}{\hbar} \frac{2x_fx_i}{T}\right)\right\}.
\label{transprob}
\end{eqnarray}

The weak trajectory $x_w(t)$ can be obtained by solving the equations (\ref{EhrenfestDoubleMome}) and (\ref{EhrenfestDoublePosi}) with the boundary values determined by the pre- and post-selections.
Alternatively, one may also obtain $x_w(t)$ directly from the amplitude (\ref{transamp}) as
\begin{eqnarray}
x_w(t)
&=&\frac{\langle x_f|U(T-t)\,x\,U(t)|\phi\rangle}{\langle x_f|U(T)|\phi\rangle}\nonumber\\
&=&\frac{\langle x_f|U(T)|x_i\rangle x_w^+(t)+\langle x_f|U(T)|-x_i\rangle x^-_w(t)}{\langle x_f|U(T)|x_i\rangle+\langle x_f|U(T)|-x_i\rangle}\nonumber\\
&=&{x_f}\frac{t}{T} - ix_i\tan \left(\frac{m}{\hbar}\frac{{x_f} {x_i}}{T}\right) \left(1 - \frac{t}{T}\right),
\label{DoubleWeak}
\end{eqnarray}
where we have used the fact that the weak trajectories for the non-superposed selections are given by the corresponding classical ones,
\begin{eqnarray}
x^\pm_w(t)
&=&\frac{\langle x_f|U(T-t)\,x\,U(t)|\pm x_i\rangle}{\langle x_f|U(T)|\pm x_i\rangle}\nonumber\\
&=&\frac{(x_f\mp x_i)t\pm x_i T}{T},
\label{wtxpm}
\end{eqnarray}
thanks to the equality (\ref{wtsingle}).

We thus find that, unlike the previous case where both the pre- and post-selections are made by position eingenstates, in the present double slit case the weak trajectory becomes complex in general, starting with the pure imaginary value
$x_w(0) = -ix_i\tan \left(\frac{m}{\hbar}\frac{{x_f} {x_i}}{T}\right)$ and ending with the real value $x_w(T) = x_f$ (see FIG.\,\ref{InterferenceDiagonal}).  This already shows that the weak trajectory does not admit the simple classical picture that we might hope for.  Hence, 
to consider its physical meaning, 
we need to examine the profile of the trajectory in the complex plane during the period $[0, T]$ closely, which is important if we are to argue the local reality of the particle in the period based on the weak value.

\begin{figure}[t]
  \begin{center}
    \includegraphics[clip,width=7.0cm]{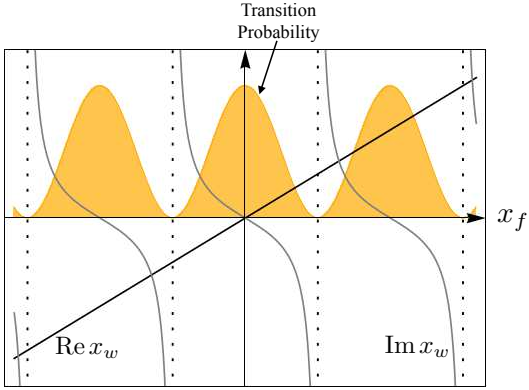}
  \end{center}
    \caption{The weak value $x_w(t)$ as a function of the post-selection $x_f$ for a fixed $t$ for $0 \le t < T$.  The thick line represents $\hbox{Re}\,x_w$ while the thin line represents $\hbox{Im}\,x_w$.
    The imaginary part $\hbox{Im}\,x_w$ diverges at the locations where the transition probability, indicated by the orange filled line, vanishes.}
  \label{fig:DoubleXfRe}
\end{figure}

To this end, let us inspect the real and imaginary parts of the weak trajectory $x_w(t)$ separately.   We then observe that the real part $\textrm{Re} \, x_w(t)$ follows just the mid-path or the average of the classical trajectories, one from $x_i$ to $x_f$ and the other from $-x_i$ to $x_f$.  
As for the imaginary part $\textrm{Im} \, x_w(t)$, we notice that it oscillates quite wildly in such a way that 
it vanishes when the interference at the screen becomes constructive while it diverges when it is destructive (see FIG.\,\ref{fig:DoubleXfRe}), as can be easily seen by comparing it with the transition probability (\ref{transprob}).
As such, the imaginary part may be regarded as an indicator of the interference effect, which can be shown to be valid in a more general context \cite{Mori}.  

The validity of our observation on the weak trajectories made for the simple two cases must further be examined by cases where more general selections are considered.  We now do this for the triple slit case, before going on to the multiple slit case later.

\section{The Triple Slit Experiment}
\begin{figure}[t]
  \begin{center}
    \includegraphics[clip,width=7.0cm]{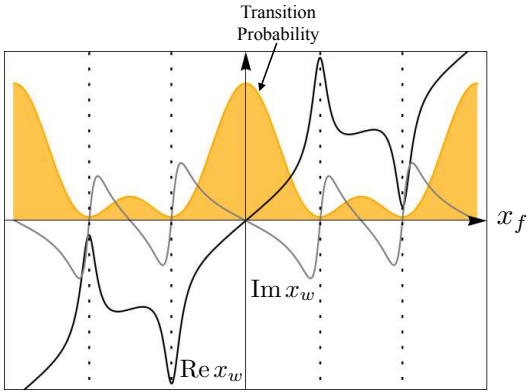}
  \end{center}
\caption{The weak value $x_w(t)$ as a function of the post-selection $x_f$ for a fixed $t$ for $0 \le t < T$.  Both $\hbox{Re}\,x_w$ (thick line) and $\hbox{Im}\,x_w$ (thin line) show 
distinctively different behaviors from the double slit case, yielding finite peaks at the locations where the transition probability becomes minimal. }
\label{fig:TripleXfRe}
\end{figure}

As a next step toward generalization, we discuss the triple slit experiment where the slits are distanced equally from each other. 
This is realized by choosing the pre-selected state as 
\begin{eqnarray}
|\phi\rangle=\frac{1}{\sqrt{3}}\left(|x_i\rangle+|0\rangle+|-x_i\rangle\right),
\end{eqnarray}
while keeping the post-selected state $|x_f\rangle$ as before. 
Assuming again the free Hamiltonian, 
we find the transition probability,
\begin{eqnarray}
&&\vert \langle x_f|U(T) \vert \phi \rangle\vert^2 
= {{m}\over {6\pi \hbar T}}
\biggl\{3+2\cos\left(\frac{m}{\hbar}\frac{2x_fx_i}{T}\right)\nonumber\\
&&\qquad\qquad +\,\, 4 \cos\left(\frac{m}{\hbar}\frac{x_fx_i}{T}\right)\cos\left(\frac{m}{\hbar}\frac{x_i^2}{2T}\right)\biggr\}.
\label{transprobthree}
\end{eqnarray}
Note that the transition probability oscillates as a function of $x_f$ on the screen, but unlike the previous double slit case it does not necessarily vanish even at the most destructive interference points (see FIG.\,\ref{fig:TripleXfRe}).  

Now, the weak trajectory $x_w(t)$ can be obtained in an analogous manner as in the double slit case, and the result is
\begin{eqnarray}
x_w(t)&=&\frac{\langle x_f|U(T-t)\,x\,U(t)|\phi\rangle}{\langle x_f|U(T)|\phi\rangle}\nonumber\\
&=&{x_f}\frac{t}{T} + g(x_i, x_f) \left(1 - \frac{t}{T}\right),
\end{eqnarray}
where $g = g(x_i, x_f)$ is a complex coefficient function given by
\begin{eqnarray}
\textrm{Re} \, g
&=& \frac{2 x_i\sin\left(\frac{m}{\hbar}\frac{x_f x_i}{T}\right)\sin\left(\frac{m}{\hbar}\frac{x_i^2}{2T}\right)}{3+2\cos\left(\frac{m}{\hbar}\frac{2x_fx_i}{T}\right)+4\cos\left(\frac{m}{\hbar}\frac{x_fx_i}{T}\right)\cos\left(\frac{m}{\hbar}\frac{x_i^2}{2T}\right)},
\nonumber\\
\textrm{Im} \, g
&=& -\frac{2 x_i\left\{2\cos\left(\frac{m}{\hbar}\frac{x_f x_i}{T}\right)+\cos\left(\frac{m}{\hbar}\frac{x_i^2}{2T}\right)\right\}\sin\left(\frac{m}{\hbar}\frac{x_fx_i}{T}\right)}{3+2\cos\left(\frac{m}{\hbar}\frac{2x_fx_i}{T}\right)+4\cos\left(\frac{m}{\hbar}\frac{x_fx_i}{T}\right)\cos\left(\frac{m}{\hbar}\frac{x_i^2}{2T}\right)},
\nonumber\\
\end{eqnarray}
for the real and imaginary parts, respectively.

Since the denominator of the function $g$ is proportional to the transition probability (\ref{transprobthree}),
and since $g$ has both  real and imaginary parts, the weak value $x_w(t)$ moves away from the origin in the complex plane
when the interference becomes destructive.   This implies that the imaginary part of the weak value $\textrm{Im} \,x_w$
continues to possess the basic property as an indicator of the interference, although it does not exhibit a simple behavior as it does in the previous case  (see FIG.\,\ref{fig:TripleXfRe}) including the nondivergence at the destructive points on account of the nonvanishing transition probability (\ref{transprobthree}).

We also notice that, unlike the previous double slit case, 
the real part of the weak value $\textrm{Re} \,x_w$ shows a synchronous behavior with the imaginary part as they share the same denominator.  
As a result, it does not allow the simple intuitive picture of the average $\frac{x_f}{T}t$ of the three classical trajectories starting
from $x = x_i,$ $0$, and $ -x_i$.
However, a notable feature still remains, that is, as a complex function it is linear in time, and hence the trajectory is a straight line connecting 
$x_w(0) = g(x_i, x_f)$ and $x_w(T) = x_f$.  In fact, it can be seen that the weak trajectory $x_w(t)$ yields an average
path under an extended notion of probability for the processes, which we shall discuss in the following sections.

\section{The Multiple Slit Experiment and the Reality Condition}

Now we consider the general case where there are $N$ slits at $x = x_1, x_2, \ldots, x_N$.  Here we have
the corresponding pre-selected state,
\begin{eqnarray}
|\phi\rangle=\sum_{n=1}^Nc_n|x_n\rangle, \hspace{5mm} c_n\in \mathbb{C},
\end{eqnarray}
and the post-selected state as $|x_f\rangle$.  
The weak trajectory can then be written as
\begin{eqnarray}
x_w(t)&=&\frac{\langle x_f|U(T-t)\,x\,U(t)|\phi\rangle}{\langle x_f|U(T)|\phi\rangle}
\nonumber\\&=&
\sum_{n=1}^N \omega_n x^{n}_w(t),
\label{nwt}
\end{eqnarray}
where we have introduced the weight factor,
\begin{eqnarray}
\omega_n=\frac{c_n\langle x_f|U(T)|x_n\rangle}{\sum_n c_n\langle x_f|U(T)|x_n\rangle}.
\label{omedef}
\end{eqnarray}
In (\ref{nwt}), the function $x^{n}_w(t)$ is just the weak trajectory with the pre-selected state $|x_n\rangle$, {\it i.e.}, 
\begin{eqnarray}
x^n_w(t) =\frac{\langle x_f|U(T-t)\,x\,U(t)| x_n\rangle}{\langle x_f|U(T)| x_n\rangle},
\label{wtn}
\end{eqnarray}
which, as we have seen in (\ref{wtsingle}), is equal to the classical trajectory $x^n_w(t) = x^n_{cl}(t)$ given by
\begin{eqnarray}
x^n_{cl}(t) :=\frac{(x_f - x_n)t +  x_nT}{T}.
\label{wtncl}
\end{eqnarray}
Plugging this expression into (\ref{nwt}), we obtain
\begin{eqnarray}
x_w(t) = x_f \frac{t}{T}+ \left(1 - \frac{t}{T}\right)\sum_{n=1}^N\omega_n x_n .
\label{MultiLinear}
\end{eqnarray}
At this point, we note that the weight factor $\omega_n$ is, in general, complex, but satisfies $\sum_n \omega_n=1$. 
Thus, the expression (\ref{nwt}) alludes to the interpretation that the weak trajectory $x_w(t)$ represents an 
average of $N$ classical trajectories going from $x_n$ over to $x_f$ with the ``complex probability'' $\omega_n$.
In fact, this simple interpretation is seen to be valid for more general cases, and can be regarded as a basic and universal
property of the weak trajectory. 

Now, let us consider the special situation in which the trajectory $x_w(t)$ becomes entirely real, Im$[x_w(t)]=0$.  This occurs if
\begin{eqnarray}
\textrm{Im}\sum_{n=1}^N \omega_n x_n 
=0,
\label{NCondition}
\end{eqnarray}
which imposes a condition on the combination of the transition amplitudes and the form of the pre-selected state.
Since varying the final point $x_f$ alters each of the transition amplitudes associated with the slits at $x_1, x_2 \ldots, x_N$, it is clear that
there are infinitely many isolated points $x_f$ for which the above condition (\ref{NCondition}) is satisfied, ensuring that the real trajectories appear there.

In fact, the condition (\ref{NCondition}) can be made simpler in terms of transition functions as
\begin{eqnarray}
\frac{d}{d x_f} {|\langle x_f|U(T)|\phi\rangle|^2} = 0.
\label{GeneralCondition5}
\end{eqnarray}
To see this, we first observe that
\begin{eqnarray}
\sum_{n=1}^N \omega_n x_n 
&=& \sum_{n=1}^N c_n \frac{\langle x_f|U(T)|x_n\rangle}{\langle x_f|U(T)|\phi\rangle}x_n
\nonumber\\
&=& \sum_{n=1}^N c_n \frac{\langle x_f|U(T)x|x_n\rangle}{\langle x_f|U(T)|\phi\rangle}
\nonumber\\
&=& \frac{\langle x_f|\left(x-\frac{p}{m}T\right)U(T)|\phi\rangle}{\langle x_f|U(T)|\phi\rangle}
\nonumber\\
&=&x_f -\frac{T}{m}\frac{\langle x_f|{p}\,U(T)|\phi\rangle}{\langle x_f|U(T)|\phi\rangle}
\nonumber\\
&=& x_f + i \hbar\frac{T}{m} \frac{d}{d x_f} \ln{\langle x_f|U(T)|\phi\rangle},
\label{imptrel}
\end{eqnarray}
where in the last equality we have used $\langle x_f|{p} = -i \hbar \frac{d}{d x_f} \langle x_f|$.
Then, taking the imaginary part of the above, we obtain
\begin{eqnarray}
\hbox{Im} \sum_{n=1}^N \omega_n x_n 
= \frac{\hbar T}{2m}\frac{d}{d x_f} \ln{|\langle x_f|U(T)|\phi\rangle|^2},
\label{EquivalenceGeneral}
\end{eqnarray}
from which the equivalence of the conditions (\ref{NCondition}) and (\ref{GeneralCondition5}) follows.  
From the latter, we can state that the weak trajectory 
$x_w(t)$ becomes entirely real at the points on the screen where the transition probability becomes stable under the change of  position.
In other words, it occurs when the profile of interference is extremal.

At this point we also mention that, in the present free case $V(x) = 0$, the weak value of the momentum $p_w$ admits the form
\begin{eqnarray}
p_w(t)&=&\frac{\langle x_f|U(T-t)\,p\,U(t)|\phi\rangle}{\langle x_f|U(T)|\phi\rangle}=\frac{\langle x_f|p\,U(T)|\phi\rangle}{\langle x_f|U(T)|\phi\rangle}\nonumber\\
&=&-i\hbar\frac{d}{dx_f}\ln \langle x_f|U(T)|\phi\rangle.
\label{GeneralMome}
\end{eqnarray}
Thus, the reality condition (\ref{GeneralCondition5}) is nothing but the reality condition $\textrm{Im}\, p_w = 0$ of the weak momentum as well.
Moreover, plugging the relation (\ref{imptrel}) into (\ref{MultiLinear}), we arrive at
\begin{eqnarray}
x_w(t) = x_f-i\hbar\left(\frac{t-T}{m}\right)\frac{d}{dx_f}\ln\langle x_f|U(T)|\phi\rangle.
\label{DoubleLinear}
\end{eqnarray}
This establishes a direct relation between the weak trajectory and the transition amplitude in the free particle case.

In the presence of the potential $V(x)$, Eq.\,(\ref{Ehrenfest}) becomes 
\begin{eqnarray}
\frac{d}{dt}p_w(t)
&=&-\left( \frac{\partial V}{\partial x}\right)_w(t),
\nonumber\\
\frac{d}{dt}x_w(t)
&=&\frac{1}{m} p_w(t).
\label{EhrenfestGeneral2}
\end{eqnarray}
From (\ref{EhrenfestGeneral2}) it is clear that, if $x_w(t)$ is purely real at any time $t$, then $p_w(t)$ is also purely real in the entire period $[0, T]$.  Taking its contraposition, one sees that, if $p_w(t)$ becomes imaginary at some $t$, then $x_w(t)$ also develops a region where it has an imaginary part.   In this sense, the momentum weak value $p_w(t)$ is a convenient quantity to inspect the reality of the weak trajectory, which is most readily done when the potential vanishes, for which it is only necessary to examine the boundary value (\ref{GeneralMome}) at $t = T$. 

Regarding the relation to the interference, one notes that (\ref{GeneralMome}) is valid
even in the presence of the potential $V(x)$ if restricted to $t = T$, {\it i.e.}, as an equation for $p_w(T)$.  Combining (\ref{GeneralMome}) with (\ref{EhrenfestGeneral2}), one realizes that the velocity of the weak trajectory $\frac{d}{dt}x_w(t)\vert_{t = T}$ at the moment when the particle hits the screen vanishes for those $x_f$ where the transition amplitude becomes extremal.  This offers the intuitive picture that, viewed from the flow of weak trajectories, the extremal constructive (or destructive) interference points are those toward which the trajectories of the particle gather (or from which they move away) in the imaginary direction for which the transition probability $|\langle x_f|U(T)|\phi\rangle|^2$ is concerned.

\section{Weak Trajectory and Which-Path Information}

So far, we have observed that, albeit via complex probability, the weak trajectory is related to the average of classical trajectories and, as such, it does not tell us from which slit the particle comes when it is detected at $x_f$ on the screen.  
In this section we briefly digress from our main line of argument and discuss how this problem can be removed, generalizing the idea presented in \cite{Mori}.  

In order to distinguish particles based on the slits they come from, all we need is to
furnish an extra $N$ degrees of ``spin'' freedom $\{ |n\rangle\}_{n = 1}^N$  which encodes the information of the slit. 
Indeed, equipped with a supplemental device to detect the spin furnished, we can prepare the pre-selected state by
\begin{eqnarray}
|\phi\rangle=\sum_{n=1}^Nc_n|x_n\rangle\otimes |n\rangle, \hspace{10mm}c_n\in \mathbb{C},
\label{PreQE}
\end{eqnarray}
so that the which-path information is gained by measuring the spin together with the position simultaneously.  This amounts to considering the observable
\begin{eqnarray}
x^{(n)}=x\otimes |n\rangle\langle n|,
\label{posionspin}
\end{eqnarray}
which fulfills $\sum_n x^{(n)}=x \otimes I$.  

Now, for the post-selection, we consider the state in the form
\begin{eqnarray}
|\psi\rangle=|x_f\rangle\otimes\sum_{n=1}^Ne_n |n\rangle, \hspace{10mm}e_n\in \mathbb{C}.
\label{PostQE}
\end{eqnarray}
As we can see easily, if we choose $e_n=\delta_{nl}$ for all $n$ and some $l$ in the post-selection (\ref{PostQE}), we pick up particles which only come from the slit at $x_l$; namely, we obtain the complete ``which-path information.''  This, however, necessarily destroys the interference pattern in accordance with quantum complementarity. 
If, instead, we choose $e_n=1/\sqrt{N}$, we lose the which-path information but maintain the interference pattern, as is well known in the context of the quantum eraser \cite{Scully}.   Our post-selection (\ref{PostQE}) yields an arbitrary interpolation between the two extremes.

With these preparations, the weak value of the observable $x^{(n)}$ defined in (\ref{posionspin}) reads 
\begin{eqnarray}
x^{(n)}_w(t)
&=& \frac{\langle \psi|U(T-t)\,x^{(n)}\,U(t)|\phi\rangle}{\langle \psi|U(T)|\phi\rangle} 
\nonumber\\
&=& \omega_n x_w^n(t),
\end{eqnarray}
where now the weight factor $\omega_n$ is slightly generalized from (\ref{omedef}) as
\begin{eqnarray}
\omega_n=\frac{c_ne^\ast_n\langle x_f|U(T)|x_n\rangle}{\sum_n c_n e^\ast_n \langle x_f|U(T)|x_n\rangle},
\label{weightQE}
\end{eqnarray}
which still satisfies the normalization condition $\sum_n \omega_n=1$. 
As before, on account of the equality we can replace $x_w^n(t)$ with the classical solution $x_{cl}^n(t)$ to obtain
\begin{eqnarray}
x^{(n)}_w(t)=  \omega_n x_{cl}^n(t).
\label{outcomeQE}
\end{eqnarray}
The outcome above shows that, by using the pre- and post-selections (\ref{PreQE}) and (\ref{PostQE}) under generic weight factors (\ref{weightQE}) together with the spin-tagged position operator (\ref{posionspin}), the weak trajectory of the particle can be inferred while preserving the interference pattern.  The only snag is that the outcome (\ref{outcomeQE}) is not quite $x_{cl}^n(t)$ but scaled with the factor $\omega_n$, which arises because of the need for renormalization of the particular element we picked up in the two selections.

\begin{figure}[t]
  \begin{center}
    \includegraphics[clip,width=7.0cm]{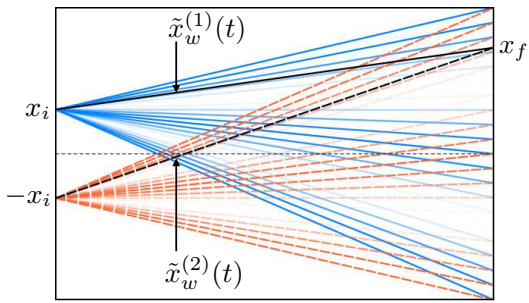}
  \end{center}
    \caption{Renormalized weak trajectories for the double slit case $N=2$ with $x_1 = x_i$, $x_2 = - x_i$.  As the post-selected state, we choose $e_n=1/\sqrt{2}$. Thus, the interference fringe reappears. 
To a given post-selection $x_f$,  if we measure $\tilde x^{(1)}_w(t)$ we find the classical path (solid line) coming from the upper slit at $x_i$, but if we measure $\tilde x^{(2)}_w(t)$ we find the classical path (dashed line) coming from the lower slit at $-x_i$.  In this measurement we can maintain the interference without conflicting with complementarity.}
  \label{fig:QuantumEraser}
\end{figure}

One may carry out this renormalization by demanding that the final position of the particle at $t= T$ be $x_f$ in accordance with the actual measurement outcome of the position.  Since $x_{cl}^n(T) = x_f$, this suggests that we consider, instead of ${x}^i_w(t)$, the renormalized weak trajectory,
\begin{eqnarray}
\tilde{x}^{(n)}_w(t):=\frac{x^{(n)}_w(t)}{\omega_n}=x_{cl}^n(t),
\label{Normalized}
\end{eqnarray}
which meets the demand $\tilde x_{w}^{(n)} (T) = x_f$ and 
yields precisely the classical path $x_{cl}^n(t)$.   Consequently, once the renormalized weak trajectory is adopted, we infer that each of the particles
which arrives at the screen comes from a single slit, rather than from two or more slits ``simultaneously.''   We illustrate this by the simplest case $N=2$ of the double slit experiment in FIG.\,\ref{fig:QuantumEraser}. 
This is an interesting observation which becomes available under the description of the weak value.  However, we also stress that this does not violate the complementarity of quantum mechanics, because the weak value is intrinsically statistical and hence cannot be attributed to a single event in the actual procedure of measurement.  The aforementioned statement is valid only in the sense of inference from the statistical outcome of the weak value.

In passing, we also note that our outcome that the weak trajectory is equal to the classical one--which is afforded under the free Hamiltonian (though this can be generalized to the case of quadratic potentials)--is not at all important in our argument for the inference.  The point is that the (renormalized) weak trajectory yields a definite function of time such that it can be regarded as some trajectory, and the question whether it coincides with the classical one or not is secondary.  However, once the trajectory is established, then it should be interesting to investigate the characteristic feature of the weak trajectory compared to the classical one, as has been done in \cite{Tanaka, Matzkin} for the case of selections where semiclassical approximation is valid.

\section{The General Case}

The foregoing argument can readily be extended to the general case of
an arbitrary pre-selected state,
\begin{eqnarray}
|\phi\rangle=\int dx_i\, \phi(x_i) |x_i\rangle, \hspace{5mm} \phi(x_i) = \langle x_i |\phi\rangle \in \mathbb{C}.
\label{genpre}
\end{eqnarray}
If we continue to use the position eigenstate $|x_f\rangle$ for the post-selected state of the free particle, the weak value of the observable $A$ is 
\begin{eqnarray}
A_w(t)
&=&\frac{\langle x_f|U(T-t)\, A\, U(t)|\phi\rangle}{\langle x_f|U(T)|\phi\rangle}\nonumber\\
&=&\int dx_i \,\omega(x_i) A_{w}(x_i; t),
\label{generalOb}
\end{eqnarray}
where $A_{w}(x_i; t)$ is the weak value when the pre-selection is made by the eigenstate $|\phi\rangle = |x_i\rangle$.  The coefficient
$\omega(x_i)$ is given by
\begin{eqnarray}
\omega(x_i)=\frac{\langle x_f|U(T)|x_i\rangle \phi(x_i)}{\int dx_i\, \langle x_f|U(T)|x_i\rangle \phi(x_i)},
\label{genpreweight}
\end{eqnarray}
which, in general, is a complex function and satisfies $\int dx_i\, \omega(x_i)=1$.
So far, the result (\ref{generalOb}) holds for an arbitrary potential $V(x)$. 

One obtains the weak trajectory when the observable $A$ is chosen to be the position operator $x$. 
In particular, for the free Hamiltonian, the equality (\ref{generalOb}) implies
\begin{eqnarray}
x_w(t) = \int dx_i \,\omega(x_i) x_{cl}(t),
\label{genident}
\end{eqnarray}
where $x_{cl}(t)$ is the classical trajectory (\ref{wtsingle}) going from $x_i$ to $x_f$.
As in the discrete case, one can immediately confirm that the relation analogous to (\ref{imptrel}),
\begin{eqnarray}
\int dx_i\, \omega(x_i) \,x_i =  x_f + i \hbar\frac{T}{m} \frac{d}{d x_f} \ln{\langle x_f|U(T)|\phi\rangle},
\label{imptrelgen}
\end{eqnarray}
holds just by replacing the sum over $n$ with the integral over $x_i$.  
Consequently,
the reality condition for the trajectory,
$\hbox{Im}\, x_w(t) = 0$, which in view of (\ref{genident}) is equivalent to
\begin{eqnarray}
\hbox{Im}\, \int dx_i\, \omega(x_i) \,x_i=0,
\label{CCondition}
\end{eqnarray}
again boils down to the condition (\ref{GeneralCondition5}).

One might think from the experience of the case of multiple slits that, since the post-selected state is in general a superposition of infinitely many position eigenstates, 
it is almost impossible to find a situation, except those already mentioned, for which the weak trajectory becomes purely real and may admit an intuitive classical picture.  
That this is not the case is readily seen by the example of the momentum eigenstate $|\phi\rangle =|p\rangle$, where one has
\begin{eqnarray}
\phi(x_i) = \langle x_i |p\rangle = \frac{1}{\sqrt{2\pi \hbar}} \, e^{i p x_i/\hbar}.
\end{eqnarray}
One then finds the weak values of $p_w$ and $x_w$ as
\begin{eqnarray}
p_w(t)&=&\frac{\langle x_f|U(T-t)\,p\,U(t)|p\rangle}{\langle x_f|U(T)|p\rangle}=p,\\
x_w(t)&=&\frac{\langle x_f|U(T-t)\,x\,U(t)|p\rangle}{\langle x_f|U(T)|p\rangle}=x_f+\frac{p}{m}(t-T).
\nonumber\\
\end{eqnarray}
These results are consistent with (\ref{Ehrenfest}) and certainly agree with the classical picture of a free particle moving with momentum $p$ and arriving at $x = x_f$ at time $t = T$.
The reality of these weak values is also consistent with the fact that the condition (\ref{GeneralCondition5}) is fulfilled by the choice $|\phi\rangle =|p\rangle$ as can be
confirmed easily:
\begin{eqnarray}
\frac{d}{dx_f}\left|\langle x_f|U(T)|p\rangle \right|^2
&=&
\frac{d}{dx_f}\left|e^{-i\frac{p^2}{2m\hbar}T}\frac{1}{\sqrt{2\pi \hbar}} \, e^{i p x_i/\hbar}\right|^2
\nonumber\\
&=&
\frac{d}{dx_f}\frac{1}{2\pi\hbar}
\nonumber\\
&=& 0.
\end{eqnarray}

A less trivial example is provided by the (complex) Gaussian state,
\begin{eqnarray}
|\psi\rangle= k \int dx_i \,  e^{i\alpha x_i^2+i\beta x_i}|x_i\rangle,
\end{eqnarray}
with real parameters $\alpha$, $\beta$, 
and a normalization constant $k$. 
Note that this pre-selected state is neither an eigenstate of the position nor of the momentum. 
The transition amplitude $\langle x_f|U(T)|\phi\rangle$ reads
\begin{eqnarray}
\langle x_f|U(T)|\phi\rangle= k \frac{\exp\left[i\frac{ x_f^2\alpha +x_f\beta -\frac{\hbar}{2m}T\beta^2}{1+2 \frac{\hbar}{m}T\alpha}\right]}{\sqrt{1+2\frac{\hbar}{m}T\alpha}},
\end{eqnarray}
which obviously meets the condition (\ref{GeneralCondition5}).  This ensures the real weak values for $p_w(t)$ and $x_w(t)$, which are explicitly given by
\begin{eqnarray}
&&p_w(t)=m\, \frac{x_f\alpha+\frac{\beta}{2}}{T\alpha+\frac{m}{2\hbar}},\\
&&x_w(t)=x_f-\left(1-\frac{t}{T}\right)\frac{x_f\alpha+\frac{\beta}{2}}{\alpha+\frac{m}{2\hbar T}}.
\end{eqnarray}
This result shows, in particular, that in the squeezing limit $\vert \alpha \vert \to \infty$ the weak trajectory reduces to the straight line $x_w(t) = (t/T)x_f$ going from the center $x = 0$ to $x = x_f$, as one expects.

Here we mention that the formula (\ref{genident}) is valid even in the presence of the potential $V(x)$, if only the classical trajectory $x_{cl}(t)$ in (\ref{genident}) is replaced by
the corresponding weak trajectory $x_{w}(x_i; t)$ which is obtained under the pre-selection $|\phi\rangle = |x_i\rangle$.
The point to be noted is that, although now $x_{w}(x_i; t)$ is not equal to $x_{cl}(t)$ (unless $V(x)$ is quadratic in $x$), 
it gives a definite trajectory obeying the boundary conditions dictated by the pre- and post-selections, {\it i.e.}, $x_{w}(x_i; 0) = x_i$ and $x_{w}(x_i; T) =x_f$.
Consequently, the formula (\ref{genident})
still admits the interpretation that the weak trajectory $x_{w}(t)$ is the average over such weak trajectories with definite boundary conditions weighted
by the complex probability $\omega(x_i)$.  This reminds us of Feynman's path-integral evaluation of the time-dependent position $x(t)$ in the transition from the initial state $|\phi\rangle$ given by the superposition (\ref{genpre}) ending up with the final state $|x_f\rangle$.  In fact, in that context our formula amounts to the arrangement of summation over the paths according to the separate classes of weak trajectories associated with distinct boundary conditions, rather than those of classical paths by the analogous classification as conventionally done in the WKB semiclassical approximation.  In this respect, we are naturally interested in the distinction between the classical trajectory $x_{cl}(t)$ and the weak trajectory $x_{w}(x_i; t)$, which has been studied earlier in \cite{Tanaka, Matzkin} from different viewpoints.

In our discussion so far, we have only considered the general case of the pre-selection (\ref{genpre}) while keeping the position eigenstate $\vert x_f \rangle$ for the post-selection $|\psi\rangle$, but it is straightforward to also generalize the post-selection as
\begin{eqnarray}
|\psi\rangle=\int dx_f\, \psi(x_f) |x_f\rangle, \hspace{5mm} \psi(x_f) = \langle x_f |\psi\rangle \in \mathbb{C}.
\end{eqnarray}
Then, combined with (\ref{genpreweight}), the weight factor for the general pre- and post-selections becomes
\begin{eqnarray}
\omega(x_i, x_f)=\frac{\psi^*(x_f) \langle x_f|U(T)|x_i\rangle \phi(x_i)}{\int dx_i dx_f\, \psi^*(x_f) \langle x_f|U(T)|x_i\rangle \phi(x_i)},
\label{genprepostweight}
\end{eqnarray}
which is normalized as $\int dx_i dx_f\, \omega(x_i, x_f) = 1$.
In fact, the complex probability distribution given by the weight $\omega(x_i, x_f)$ in the most general case (\ref{genprepostweight}) is precisely that encapsulated by the complex probability measure 
proposed earlier in \cite{Hosoya, Morita} as a proper measure for the weak value assigned to the quantum process specified by the pre- and post-selections.

\section{Lloyd's Mirror} 

Finally, we wish to discuss a nontrivial example in which both the pre- and post-selections are provided by position eigenstates but the system admits more than one classical trajectory.  In this case, one cannot expect the weak trajectory to agree with any of the classical ones, even though the boundary values at $t =0$ and $t = T$ are identical in the two cases.   Such an example offers another type of grounds for examining the physical significance of the weak value in addition to those considered so far.

\begin{figure}[t]
\begin{center}
\includegraphics[width=7cm]{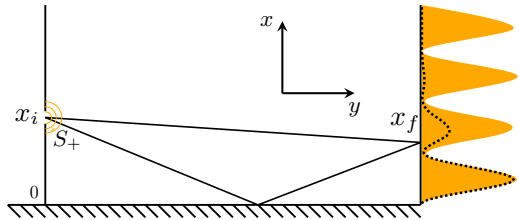}
\caption{Lloyd's mirror experiment in which an interference pattern appears with a single hole.  
The orange filled curve describes the transition probability in our simplified situation where the slit $S_+$ is point-like, whereas the dotted curve describes the transition probability in a more realistic case where the slit $S_+$ is finite in size and the particle has a Gaussian distribution. 
}
\label{QuantumInterferenceLloyd}
\end{center}
\end{figure}

The system we consider is a free particle confined in a half plane, where we have an infinite potential wall at $x=0$, {\it i.e.},  
$V(x)=\infty$ for $x<0$ (see FIG.\,\ref{QuantumInterferenceLloyd}).   Classically, this implies that the particle bounces back when it
hits the wall.  In quantum mechanics, it is known that there exists a one-parameter family of possible boundary conditions at the wall (see, for example, \cite{RS}), but for simplicity we choose to work in the Dirichlet boundary condition
$\phi(0) = 0$ for the wave function $\phi(x) = \langle x | \phi \rangle$.   

Now, let our pre- and post-selections be given by $| \phi \rangle = | x_i \rangle$ and $| \psi \rangle = | x_f \rangle$, respectively, which corresponds to the situation in which the particle
departs from the position $x = x_i$ and ends up with the position $x = x_f$ during the period $[0, T]$. 
In particular, we fix the initial position of the particle at $x_i$ and monitor the weak trajectories for various final positions $x_f$.
As in the double slit experiment, we have two distinct classes of available paths, one that goes directly to the point of post-selection $x_f$ and the other that hits the wall before arriving at $x_f$, which generate an interference pattern on the screen.  This is known as Lloyd's mirror experiment in which interference can be observed even with a single hole (light source) device.

Summing over all possible paths belonging to the two distinct classes mentioned above, and taking the boundary condition at the wall into account, one finds that the Feynman kernel for this case reads \cite{FT}
\begin{eqnarray}
\langle x_f|U(T)|x_i\rangle
= \sqrt{\frac{m}{2\pi i \hbar T}}
\left(
 e^{\frac{i m(x_f -x_i)^2}{2\hbar T}}
-e^{\frac{i m(x_f+x_i)^2}{2\hbar T}}
\right).
\nonumber\\
\label{fkl}
\end{eqnarray}
The transition probability is then found to be
\begin{eqnarray}
\left|\langle x_f|U(T)|x_i\rangle\right|^2
=\frac{2m}{\pi \hbar T}\sin^2\left(\frac{m}{\hbar}\frac{x_f x_i}{T}\right),
\end{eqnarray}
which yields a sinusoidal interference pattern analogous to that observed in the double slit case.

The weak trajectory is now given by
\begin{eqnarray}
x_w(t)&=&\frac{\langle x_f|U(T-t)\,x\,U(t)|x_i\rangle}{\langle x_f|U(T)|x_i\rangle}
\nonumber\\
&=&\frac{\int_0^\infty dx\, \langle x_f|U(T-t) |x\rangle x \langle x | U(t)|x_i\rangle}{\langle x_f|U(T)|x_i\rangle}.
\end{eqnarray}
Using the expression (\ref{fkl}) for the kernel, one obtains
\begin{eqnarray}
&&x_w(t)\nonumber\\
&&=i \frac{e^{-i \frac{m}{\hbar}\frac{x_f x_i}{T}}(t(x_f-x_i)+T x_i)\textrm{Erfi}\left[
\frac{(t(x_f-x_i)+Tx_i)}{\sqrt{-i 
tT(T-t)}}\sqrt{\frac{m}{\hbar}}\right]}{\left(e^{i \frac{m}{\hbar}\frac{x_f x_i}{T}}-e^{-i \frac{m}{\hbar}\frac{x_f x_i}{T}}\right)T}\nonumber\\
&& +\, i \frac{e^{i \frac{m}{\hbar}\frac{x_f x_i}{T}}(t(x_f+x_i)-T x_i)\textrm{Erfi}\left[
\frac{(-t(x_f+x_i)+Tx_i)}{\sqrt{-i tT(T-t)}}\sqrt{\frac{m}{\hbar}}\right]}{\left(e^{i \frac{m}{\hbar}\frac{x_f x_i}{T}}-e^{-i \frac{m}{\hbar}\frac{x_f x_i}{T}}\right)T},\nonumber\\
\end{eqnarray}
where Erfi$[z]$ is the error function defined by 
\begin{eqnarray}
\textrm{Erfi}[z]=\frac{\textrm{Erf}[i z]}{i}=\frac{2}{i\sqrt{\pi}}\int_0^{i z}e^{-t^2}dt.
\end{eqnarray}

\begin{figure}[t]
\begin{center}
\includegraphics[width=7cm]{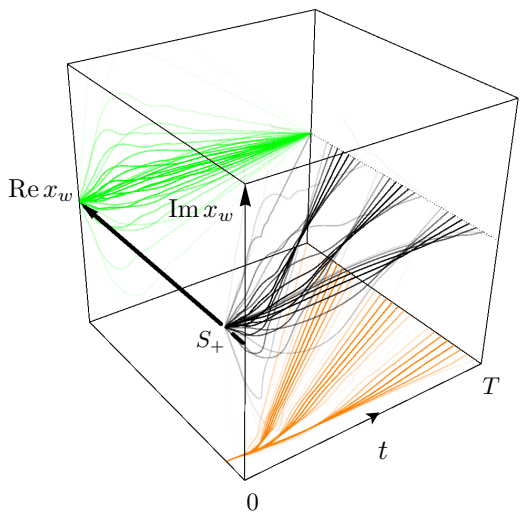}
\caption{ The weak trajectories $x_w(t)$ in the complex plane 
for a number of different post-selections $x_f$ plotted with density in proportion to the transition probability (the notations are the same as in FIG.\,\ref{InterferenceDiagonal}).
The imaginary part $\textrm{Im} \ x_w$ varies violently while the real part $\textrm{Re} \ x_w$ follows more or less a smoothed average of the two classical trajectories.  
}
\label{InterferenceDiagonalLloyd}
\end{center}
\end{figure}

Note that $x_w(t)$ is no longer linear in $t$; actually it exhibits a rather complicated behavior as a function of time as can be seen from the numerical result shown in FIG.\,\ref{InterferenceDiagonalLloyd}.

As shown in FIG.\,\ref{InterferenceDiagonalLloyd}, in general the weak value $x_w(t)$ is not real, which is also confirmed by examining  
 the reality condition $\textrm{Im}\, p_w = 0$
based on the relation (\ref{GeneralMome}).  If we adopt the same criterion (ignoring the possible problem of self-adjointness of the momentum operator $p$ on the half-line) in the present case, we find
\begin{eqnarray}
\textrm{Im} \ p_w(T)=-\frac{mx_i}{T}\cot\left(\frac{m}{\hbar}\frac{x_f x_i}{T}\right).
\end{eqnarray}
From this, we can infer that $ x_w(t)$ is also complex, even though $x_w(t)$ becomes real at the two boundaries at $t = 0$ and $t = T$. 

Despite these complications, in FIG.\,\ref{InterferenceDiagonalLloyd} we observe that the real part $\textrm{Re} \ x_w(t)$ yields more or less a smoothed trajectory of the average of the two classical trajectories.  It should be noted, however, that when the interference is destructive, even the real part tends to show a violent behavior.  In contrast, the imaginary part fluctuates almost chaotically during the period except for the ends where the trajectory is fixed by the value of the two selections.

\section{Conclusion and Discussions}

In the present paper, we have studied the time development of the weak value for the position of a particle--the weak trajectory--and thereby examined to what extent the weak value admits an intuitive picture, which is important for associating reality to the particle motion in quantum mechanics in one way or the other.  More specifically, we have defined the weak trajectory $x_w(t)$ during the period $[0, T]$ based on the time-dependent weak value (\ref{tdwv}) for the position observable $A = x$ under a given transition process specified by the pre- and post-selected states at time $t=0$ and $t = T$, respectively.   Armed with this, 
we have examined the behavior of the weak trajectory so defined in various situations, starting from the simple case where the particle is free and resides 
at a definite place both at $t=0$ and $t = T$ and ending with a completely general case of post-selections.  These include the (simplified) double slit experiment case and the triple and multiple slit cases where interference takes place even with the detection of particles, that is, where a genuine quantum effect arises.  We also considered Lloyd's mirror which allows interference with a single slit in the presence of a reflecting wall.

In the simplest case where the location of the particle is given by position eigenstates at the two ends, we have seen that our weak trajectory $x_w(t)$ is real and linear in time, and coincides precisely with the classical trajectory $x_{cl}(t)$.
This reassuring result cannot arise in the double slit case, because the pre-selection is made by a superposed state and hence the particle is not localized at a single point.  Accordingly, no unique classical solution exists, and the weak trajectory $x_w(t)$ becomes complex in this case.  This is, however, harmless because, if we focus on the real part $\textrm{Re}\, x_w(t)$, we observe that it continues to yield the average of the two classical trajectories associated with the two position eigenstates appearing in the superposition of the pre-selected state.   
This observation is no longer valid when we go over to the triple slit or more generally the multiple slit cases.   
Nonetheless, the weak trajectory $x_w(t)$ still affords, as a whole, interpretation as the 
average over the trajectories in the extended sense of  ``complex probability'' attached to the transitions associated with the position eigenstates in the superposition.
This is actually the notion of probability which is used to define the probability measure for the weak value mentioned earlier \cite{Hosoya, Morita}.

We have also noticed that it is possible to define a modified position operator such that its weak value yields the classical trajectory from an individual slit specified by the operator.  This is accomplished by supplementing the extra ``spin'' degrees of freedom and preparing the pre-selected state
in the maximally entangled state between the position and the spin following the idea of the quantum eraser \cite{Scully}.   The point is that, because of the intrinsic statistical nature of the weak value, this can be done without conflicting with the complementarity of quantum mechanics (for a fuller discussion on this issue, see \cite{Mori}).

Our results on the various pre-selections suggest that, except for a few cases such as the simplest situation where it coincides with the classical trajectory, the weak trajectory $x_w(t)$ becomes necessarily complex.  When the particle is free, the weak trajectory is shown to 
be linear in time and is related directly to the transition amplitude (\ref{DoubleLinear}).  From this, one then realizes that the
reality condition of the weak trajectory is just the extremal condition for the transition probability (\ref{GeneralCondition5}).  In other words, the weak trajectory $x_w(t)$ becomes real if and only if the interference is maximally constructive or maximally destructive, unless no interference occurs on the whole screen.   One example of the last case is provided by the momentum eigenstate $|\phi\rangle =|p\rangle$ for which both the weak trajectory 
$x_w(t)$ and the weak momentum $p_w(t)$ agree with their classical counterparts.  This example is worth noting, because the momentum eigenstate $ |p\rangle$ describes the wave nature of the constant momentum state and yet it can reveal the classical particle nature if one monitors the weak values of the relevant physical quantities.

The imaginary part $\textrm{Im}\, x_w(t)$ of the weak trajectory is also no idle quantity.   Indeed, since (\ref{DoubleLinear}) suggests that the imaginary part $\textrm{Im}\, x_w(t)$ diverges when the transition amplitude vanishes, one recognizes the significance of $\textrm{Im}\, x_w(t)$ as a quantity indicating the strength of interference, as one can easily gather from the outcomes of the double and triple slit examples (for the precise meaning of the imaginary part in the context of interference, see \cite{Mori}).

Finally, we have touched upon Lloyd's mirror as an instance of allowing interference with a single slit.  This offers an intriguing  model where the weak trajectory is intrinsically quantum, in view of the fact, even if the initial and final positions of the particle are completely specified, no unique classical trajectory is allowed as in the double slit case, because of the presence of the perfectly reflecting wall.  Our result shows that the weak trajectory is mostly close to, but not quite equal to, the average of the two possible classical trajectories.

In conclusion, our investigation on 
the weak trajectory $x_w(t)$ 
shows that it admits an intuitive picture of particle trajectory in some particular cases, but largely it is to be interpreted as an average based on the complex probability associated with the given process of transitions.  Obviously, to obtain a fuller grasp of what the weak trajectory is, we need to go beyond the essentially free particle cases presented here and investigate more general cases which involve nontrivial potentials.  
The power of the weak trajectory, or more generally, the significance of the time-dependent weak value $A_w(t)$ that can be defined for any observable $A$, will then become clearer and, presumably, be understood in line with the time-symmetric formulation of quantum mechanics \cite{ABL, Danan}.  Since the weak value is a quantity obtained under the weak limit of measurement, if, ideally, it is independent of the type of weak measurement performed, it should reveal some intrinsic aspect of the physical property peculiar to the process.   What that aspect can be will be learnt only by accumulating examples, and the elementary ones presented in this paper are hopefully of some help in moving toward this goal.

\section*{Acknowledgments}
I. T. thanks Prof. A. Wipf for useful discussions.  
This work was supported in part by the Grant-in-Aid for Scientific Research (C), No.~25400423 of MEXT, and by the Center for the Promotion of Integrated Sciences (CPIS) of Sokendai.



\begin{thebibliography}{2}
\bibitem{Aharonov}
Y. Aharonov, D. Z. Albert, and L. Vaidman, Phys. Rev. Lett. $\bold{60}$, 1531(1988).
\bibitem{Vaidman}
L. Vaidman, Found. Phys. $\bold{26}$, 895 (1996).
\bibitem{EPR}
A. Einstein, B. Podolsky, and N. Rosen, Phys. Rev. {\bf 47}, 777 (1935).
\bibitem{Kocsis}
S. Kocsis, B. Braverman, S. Ravets, M. J. Stevens, R. P. Mirin, L. K. Shalm, and A. M. Steinberg, Science $\bold{332}$, 1170 (2011).
\bibitem{Wiseman}
H.M. Wiseman, New J. Phys. $\bold{9}$, 165 (2007).
\bibitem{QuantumPotential}
C. Philippidis, C. Dewdney, and B. J. Hiley, Nuovo Cimento B $\bold{52}$,  15(1979).
\bibitem{Tanaka}
A. Tanaka, Phys. Lett. A $\bold{297}$, 307(2002).
\bibitem{Matzkin}
A. Matzkin, Phys. Rev. Lett. $\bold{109}$, 150407 (2012).
\bibitem{ABL}
Y. Aharonov, P. G. Bergmann, and L. Lebowitz, Phys. Rev. $\bold{134}$,  B1410 (1964).
\bibitem{AharonovAve}
Y. Aharonov and A. Botero, Phys Rev.  A$\bold{72}$, 052111 (2005).
\bibitem{Mori}
T. Mori and I. Tsutsui, 
arXiv:1410.0787.
\bibitem{Hosoya}
A. Hosoya and M. Koga, J. Phys. A: Math. Theor. $\bold{44}$, 415303 (2011).
\bibitem{Morita}
T. Morita, T. Sasaki, and I. Tsutsui, PTEP 053A02 (2012).
\bibitem{Scully}
M. O. Scully, B. G. Englert, and H. Walther, Nature (London) $\bold{351}$, 111(1991).
\bibitem{Dressel}
J. Dressel  and A. N. Jordan, Phys. Rev. A $\bold{85}$, 012107 (2012).
\bibitem{RS}
M. Reed and B. Simon,
{\it Fourier Analysis, Self-Adjointness}, Methods of Modern Mathematical Physics, 
(Academic Press, Orlando, FL, 1975), Vol. 2.
\bibitem{FT}
T. F\"{u}l\"{o}p and I. Tsutsui, Phys. Lett. A. $\bold{264}$, 366(2000).
\bibitem{Danan}
A. Danan, D. Farfurnik, S. Bar-Ad, and L. Vaidman, Phys. Rev. Lett. $\bold{111}$, 240402 (2013).
\end{thebibliography}
\end{document}